\documentstyle[12pt]{article}

\let\ni=\noindent

\begin{document}

\baselineskip 0.75cm

\renewcommand{\thefootnote}{\fnsymbol{footnote}}

\newcommand{\CKM}{Cabibbo--Kobayashi--Maskawa }

\newcommand{\SM}{Standard Model }

\pagestyle {plain}

\setcounter{page}{1}

\pagestyle{empty}

~~~
\hfill IFT/99-27

\vspace{0.3cm}

{\large\centerline{\bf An unorthodox alternative for righthanded neutrinos:}}

{\large\centerline{\bf lefthanded see--saw{\footnote{Supported in part by the 
Polish KBN--Grant 2 P03B 052 16 (1999--2000).}}}}

\vspace{0.8cm}

{\centerline {\sc Wojciech Kr\'{o}likowski}}

\vspace{0.8cm}

{\centerline {\it Institute of Theoretical Physics, Warsaw University}}

{\centerline {\it Ho\.{z}a 69,~~PL--00--681 Warszawa, ~Poland}}

\vspace{0.5cm}

{\centerline{\bf Abstract}}

\vspace{0.3cm}

A new lefthanded see--saw mechanism is constructed, implying both the
smallness of active--neutrino masses and decoupling of heavy passive
neutrinos, similarly to the situation in the case of conventional
see--saw. But now, in place of the conventional righthanded neutrinos,
the lefthanded sterile neutrinos play the role of heavy passive
neutrinos, the righthanded neutrinos and righthanded sterile neutrinos
being absent. Here, the lefthanded sterile neutrinos are different
from charge conjugates of conventional righthanded neutrinos because
their lepton numbers differ. In this case, the neutrino mass term is
necessarily of pure Majorana type.

\vspace{0.2cm}

\ni PACS numbers: 12.15.Ff , 14.60.Pq , 12.15.Hh .

\vspace{0.5cm}

\ni November 1999

\vfill\eject

~~~
\pagestyle {plain}

\setcounter{page}{1}

 As is well known, the popular see--saw mechanism [1] provides us with
a gentle way of introducing small neutrino masses into the original
\SM, where Dirac--%
type masses are zero due to the absence of
righthanded neutrinos. In fact, the actual righthanded neutrinos (of
all three generations) get in this mechanism large (by assumption)
Majorana--type masses and so, become practically decoupled from
lefthanded neutrinos that are allowed to carry only small
Majorana--type masses. Such large righthanded Majorana masses are, on
the other hand, related to an expected high energy scale at which
lepton number violation ought to appear.

 In this note, we present a new, {\it a priori} possible mechan\-ism,
where the role played in the see--saw mechanism by right\-handed
neutrinos is taken over by hypothetic {\it lefthanded} sterile
neutrinos free of any \SM charges. Thus, the new mechanism may be
called {\it lefthanded see--saw}. In this case, both righthanded
neutrinos and righthanded sterile neutrinos are conjectured to be
absent, $\nu_R \equiv 0 $ and $\nu^{(s)}_R \equiv 0 $, so that


\begin{equation}
\nu \equiv \nu_L \;\;,\;\;\nu^{(s)} \equiv \nu^{(s)}_L
\end{equation}


\ni are active and sterile neutrinos, respectively (for all three neutrino 
generations).

 At first sight it may seem that it is sufficient to reinterpret the
sterile neutrino $\nu^{(s)}_L $ as the charge conjugate $(\nu_R)^c $
of conventional righthanded neutrino $\nu_R$ in order to return to the
safe ground of familiar see--saw. But such a reinterpretation is
impossible if $\nu^{(s)}_L $, like $ \nu_L $, has the lepton number $
L = +1 $ (this would be also excluded if its $ L $ were 0). In this
argument $ L $ is assumed to be well defined, even when it is
violated.

 In the new situation, the neutrino mass term is of pure Majorana type


\begin{eqnarray}
-{\cal L}_{\rm mass} & = &
\frac{1}{2}\left(\overline{(\nu_L)^c}\,,\,\overline{ (\nu_L^{(s)})^c}
\right) \left(\begin{array}{cc} m^{(L)} & \mu^{(L)} \\ \mu^{(L)} &
m^{(L)}_s \end{array} \right) \left(\begin{array}{c} \nu^{L} \\
\nu^{(s)}_L \end{array} \right) + {\rm h.c.} \nonumber \\ & = &
\frac{1}{2} \left(\overline{(\nu_M)^c}\,,\,\overline{(\nu_M^{(s)})^c}
\right) \left(
\begin{array}{cc} m^{(L)} & \mu^{(L)} \\ \mu^{(L)} & m^{(L)}_s \end{array} 
\right) \left(\begin{array}{c} \nu^{M} \\ \nu^{(s)}_M \end{array} \right) \;,
\end{eqnarray}


\ni where


\begin{equation}
\nu_M \equiv \nu_L + (\nu_L)^c\;\;,\;\; \nu^{(s)}_M \equiv \nu^{(s)}_L +
(\nu^{(s)}_L)^c
\end{equation}

\ni (in the case of one neutrino generation). In the case of three
neutrino generations, real numbers $ m^{(L)} $, $ m^{(L)}_s $ and $
\mu^{(L)} $ become $ 3\times 3 $ matrices $ \widehat{m}^{(L)}_s $, $
\widehat{m}^{(L)}_s $ and $ \widehat{\mu}^{(L)}$, real and symmetric
for simplicity, while neutrino fields $ \nu_L $ and $ \nu^{(s)}_L $
transit into the field columns $ \vec{\nu }_L =
\left(\nu_{e\,L}\,,\,\nu_{\mu\,L}\,,\,\nu_{\tau\,L} \right)^T $ and $
\vec{\nu}^{(s)}_L =
\left(\nu^{(s)}_{e\,L}\,,\,\nu^{(s)}_{\mu\,L}\,,\,\nu^{(s) }_{\tau\,L}
\right)^T $.

 In Eq. (2), all four terms violate lepton number ($\Delta L = \pm 2 $), the 
terms proportional to $\mu^{(L)}$ and $ m^{(L)}$ break electroweak symmetry, 
while the term proportional to $m_s^{(L)}$ is electroweak gauge invariant. The 
terms $\mu^{(L)}$ and $ m^{(L)}$ may be generated spontaneously by Higgs 
mechanism, the first --- by the linear Higgs coupling (of the Majorana type):


\begin{equation}
{\cal L}_H = \frac{1}{2} g_H \left[ \overline{(l_L)^c} H \nu^{(s)}_L - 
\overline{l_L} H^c (\nu^{(s)}_L)^c \right] + 
{\rm h.c.}\;\;,\;\;\mu^{(L)} \equiv g_H \langle H^\circ \rangle
\end{equation}

\ni and the second --- by the familiar bilinear effective Higgs
coupling (also of the Majorana type):


\begin{equation}
{\cal L}_{HH} = \frac{1}{4 M} g_{HH} \left[\overline{(l_L)^c}\vec{\tau} l_L 
\right] \cdot \left( H^T i \tau_2 \vec{\tau} H\right) + {\rm h.c.}\;\;,\;\;
m^{(L)} \equiv \frac{1}{M} g_{HH} \langle H^\circ \rangle^2
\end{equation}

\ni (see {\it e.g.} Ref. [2]). Here, $\overline{(l_L)^c} = (l_L)^T C^{-1} i 
\tau_2 $ and


\begin{eqnarray}
l_L = \left(\begin{array}{c} \nu_L \\ l^-_L \end{array} \right) & , & H = 
\left(\begin{array}{c} H^+ \\ H^\circ \end{array}\right)\;, \nonumber \\ 
(l_L)^c = i \tau_2 \left( \begin{array}{c} (\nu_L)^c \\ (l^-_L)^c \end{array}
\right) = \left( \begin{array}{c} (l^-_L)^c \\ - (\nu_L)^c \end{array} \right)
& , & H^c = i \tau_2 \left( \begin{array}{c} H^{+\,c} \\ H^{\circ\,c} 
\end{array}\right) = \left( \begin{array}{c} H^{\circ\,c} \\ -H^{+\,c}
\end{array}\right)\;,
\end{eqnarray}

\ni with $(\nu_L)^c = C\overline{\nu}_L^T $, $(l^-_L)^c =
C\overline{l^{-\,T}_L }$ and $H^{+\,c} = H^{+\,\dagger} = H^-$,
$H^{\circ\,c} = H^{\circ\,\dagger}$ ($H^{c\,\dagger} = - H^{T} i
\tau_2 $). In Eq. (5), $ M $ is a large mass scale probably related to
the GUT scale, so that the inequality $ m^{(L)} \ll \mu^{(L)}$ is
plausible. Note that the lefthanded sterile neutrino $\nu_L^{(s)} $, a
\SM scalar, may be an SU(5) scalar. However, it cannot be an SO(10)
covar%
iant (say, a scalar or a member of 16--plet), since the SO(10)
formula $ Y = 2 I^{(R)}_3 + B - L $ for weak hypercharge does not work
in the case of $\nu_L^{ (s)}$ with $ L = 1 $ and $ B = 0 $ ($ Y = 0 $
and $ I^{(R)}_3 = 0 $ imply $ B - L = 0 $). Thus, the existence of
$\nu_L^{(s)}$ breaks dynamically the SO(10) symmetry, unless
$\nu_L^{(s)}$ gets $ B - L = 0 $ ({\it e.g.} $ L = 0 = B $ or $ L = 1
= B $), when it may be an SO(10) scalar (but, in absence of $ \nu_R $,
the 16--plet is still not completed). If $ L = 0 = B $, then of the
intrinsic quantum numbers only spin 1/2 and chirality --1 are carried
by $\nu^{ (s)}_L $.

 After its diagonalization, the mass term (2) becomes


\begin{equation}
-{\cal L}_{\rm mass} =
\frac{1}{2}\left(\overline{\nu}_I\,,\,\overline{ \nu}_{II} \right)
\left(\begin{array}{cc} m_I & 0 \\ 0 & m_{II} \end{array} \right)
\left( \begin{array}{c} \nu_I \\ \nu_{II} \end{array} \right) \;,
\end{equation}

\ni where 


\begin{eqnarray}
\nu_{I} & = & \nu_M \cos \theta - \nu^{(s)}_M \sin \theta\;, \nonumber \\ 
\nu_{II} & = & \nu_M \sin \theta + \nu^{(s)}_M \cos \theta
\end{eqnarray}

\ni with $\tan \theta = (m_{II} - m^{(L)}_s)/\mu^{(L)} $, and


\begin{equation}
m_{I,\,II} = \frac{m^{(L)} + m^{(L)}_s}{2} \mp \sqrt{\left( \frac{m^{(L)} - 
m^{(L)}_s}{2}\right)^2 + \mu^{(L)\,2}}\;.
\end{equation}

\ni Assuming that


\begin{equation}
0 \leq m^{(L)} \ll \mu^{(L)} \ll m^{(L)}_s\;,
\end{equation}


\ni we obtain the neutrino mass eigenstates 


\begin{equation}
\nu_I \simeq \nu_M - \frac{\mu^{(L)}}{m^{(L)}_s} \nu^{(s)}_M \simeq
\nu_M \;,\; \nu_{II} \simeq \nu^{(s)}_M + \frac{\mu^{(L)}}{m^{(L)}_s}
\nu_M \simeq \nu_M^{(s)}
\end{equation}


\ni related to the neutrino masses


\begin{equation}
m_I \simeq  - \frac{\mu^{(L)\,2}}{m^{(L)}_s}\;,\;m_{II} \simeq m^{(L)}_s\;.
\end{equation}


\ni Thus, $|m_I| \ll m_{II}$, what practically decouples $\nu_L^{(s)}$
from $\nu_L $. Here, the minus sign at $ m_I $ is evidently irrevelant
for $ \nu_I $ which, as a relativistic particle, is kinematically
characterized by $ m^2_I $.

 New Eqs. (10) --- (12) give us a purely lefthanded counterpart of the popular 
see--saw mechanism, where the assumption 


\begin{equation}
0 \leq m^{(L)} \ll m^{(D)} \ll m^{(R)}
\end{equation}


\ni implies the neutrino mass eigenstates


\begin{equation}
\nu_I \simeq \nu_M \equiv \nu_L +(\nu_L)^c \;,\;\nu_{II} \simeq \nu'_M \equiv
\nu_R + (\nu_R)^c
\end{equation}


\ni connected with the neutrino masses 


\begin{equation}
m_I \simeq  - \frac{m^{(D)\,2}}{m^{(R)}} \;,\; m_{II} \simeq m^{(R)}\;.
\end{equation}

\ni In that case, the neutrino mass term has the form

\begin{eqnarray}
-{\cal L}^{\rm conv}_{\rm mass} & = & \frac{1}{2}\left(
\overline{(\nu_L)^c} \,,\,\overline{\nu_R} \right)
\left(\begin{array}{cc} m^{(L)} & m^{(D)} \\ m^{ (D)} & m^{(R)}
\end{array} \right) \left( \begin{array}{c} \nu_L \\ (\nu_R)^c
\end{array} \right) + {\rm h.c.} \nonumber \\ & = & \frac{1}{2}
\left( \overline{\nu_M}\,,\,\overline{\nu'_M} \right) \left(
\begin{array}{cc} m^{(L)} & m^{(D)} \\ m^{(D)} & m^{(R)} \end{array}
\right) \left(\begin{array}{ c} \nu_M \\ \nu'_M \end{array} \right)
\;.
\end{eqnarray}


\ni So, from Eq. (15) $|m_I| \ll m_{II}$, what practically leads to
decoupling $\nu_R $ from $\nu_L $. But, while in Eq. (15) the
magnitude of neutrino Dirac mass $ m^{(D)}$ may be compared with the
mass of corresponding charged lepton, in Eq. (12) the magnitude of
$\mu^{(L)}$, responsible for the coupling $(1/2)
\mu^{(L)}\left[\overline{(\nu_L)^c} \nu_L^{(s)} +
\overline{\nu_L}(\nu_L^{(s)} )^c \right] $ + h.c., may be quite
different (perhaps smaller).

 In the general case of three neutrino generations, Eqs. (11) and (12) are 
replaced by


\begin{equation} 
\vec{\nu}_I \simeq \vec{\nu}_M \equiv \vec{\nu}_L + (\vec{\nu}_L)^c
\;,\; \vec{\nu}_{II} \simeq \vec{\nu}^{(s)}_M \equiv \vec{\nu}_L^{(s)}
+ (\vec{\nu }_L^{(s)})^c
\end{equation}

\ni and

\begin{equation}
\widehat{m}_I \simeq - \widehat{\mu}^{(L)}\left(\widehat{m}^{(L)}_s
\right)^{ -1} \widehat{\mu}^{(L)}\;,\;\widehat{m}_{II} \simeq
\widehat{m}^{(L)}_s
\end{equation}

\ni (compare {\it e.g.} Ref. [3]). Here, $\vec{\nu}_L =
(\nu_{\alpha\,L})$, $\vec{\nu}_I = (\nu_{I\,\alpha})$, $\widehat{m}_I
= (m_{I\,\alpha \beta})$, {\it etc.} ($\alpha\,,\,\beta =
e\,,\,\mu\,,\,\tau $). The Hermitian mass matrices $\widehat{m}_I $
and $\widehat{m}_{II}$ ($\widehat{m}^{(L)}$, $\widehat{m}^{(L)}_s $
and $\widehat{\mu}^{(L)}$ were taken real and symmetric for
simplicity) can be diagonalized with the use of unitary matrices
$\widehat{U}_I = \left( U_{I\,\alpha\,i}\right)$ and $\widehat{U}_{II}
= \left( U_{II\,\alpha\,i}\right)$, respectively, giving the neutrino
mass eigenstates


\begin{eqnarray}
\nu_{I\,i} & = & \sum_\alpha\left(\widehat{U}_{I}^\dagger \right)_{i\,\alpha}
\nu_{I\,\alpha}  = \sum_\alpha U^*_{I\,\alpha\,i} \nu_{I\,\alpha}\;, \nonumber
\\ \nu_{II\,i} & = & \sum_\alpha\left(\widehat{U}_{II}^\dagger \right)_{i\,
\alpha} \nu_{II\,\alpha}  = \sum_\alpha U^*_{II\,\alpha\,i} \nu_{II\,\alpha}
\end{eqnarray}

\ni and the corresponding neutrino masses

\begin{eqnarray}
\delta_{ij} m_{I\,i} & = & \sum_{\alpha \beta}
\left(\widehat{U}_{I}^\dagger \right)_{i\,\alpha} m_{I\,\alpha
\beta}\left(\widehat{U}_{I} \right)_{\beta\, j} = \sum_{\alpha \beta}
U^*_{I\,\alpha\,i} U_{I\,\beta\,j} m_{I\,\alpha \beta}\;, \nonumber \\
\delta_{ij} m_{II\,i} & = & \sum_{\alpha \beta} \left(
\widehat{U}_{II}^\dagger \right)_{i\,\alpha} m_{II\,\alpha \beta}
\left( \widehat{U}_{II} \right)_{\beta\,j} = \sum_{\alpha \beta}
U^*_{II\,\alpha\,i} U_{II\,\beta\,j} m_{II\,\alpha \beta}
\end{eqnarray}

\ni ($i,j = 1,2,3$). In our case, $ U_{I \alpha i}$ and $ U_{II \alpha
i}$ are real, while $ m_{I\,\alpha \beta} \simeq -\sum_{\gamma \delta}
\mu^{(L)}_{\alpha \gamma}$ $(\widehat{m}^{(L)\,-1}_s)_{\gamma
\delta}\mu^{(L) }_{\delta \beta}$ and $ m_{II\,\alpha \beta} \simeq
m^{(L)}_{s \alpha \beta}$ due to Eqs. (18). When deriving Eq. (19), we
assume that in the original lepton Lagrangian the charged--lepton mass
matrix is diagonal and so, its diagonalizing unitary matrix
$\widehat{U}^{(l)}$ is trivially equal to the unit matrix. Then, the
lepton counterpart of the \CKM matrix $\widehat{V} =\widehat{
U}^{(\nu)\,\dagger}\widehat{U}^{(l)}$ becomes equal to
$\widehat{U}^{(\nu)\, \dagger} = \widehat{U}^{\dagger}_I $, thus $V_{i
\alpha} = U^*_{I \alpha i}$.

 From Eqs. (17) and (19) we conclude for neutrino fields that 

\begin{eqnarray}
\nu_{\alpha\,L} & \simeq & \left( \nu_{I\,\alpha} \right)_L = \sum_i 
U_{I\,\alpha\,i} \left(\nu_{I\,i} \right)_L \;, \nonumber \\
\nu_{\alpha\,L}^{(s)} & \simeq & \left( \nu_{II\,\alpha} \right)_L = \sum_i 
U_{II\,\alpha\,i} \left(\nu_{II\,i} \right)_L \;.
\end{eqnarray}

\ni Thus, the oscillation probabilities (on the energy shell) for active--%
neutrino states read (in the vacuum):

\begin{equation}
P(\nu_\alpha \rightarrow \nu_\beta) = |\langle \nu_\beta| e^{iPL}|
\nu_\alpha\rangle|^2 \simeq \delta_{\alpha \beta} - 4 \sum_{i<j} U^*_{I \beta 
j} U_{I \alpha j}U_{I \beta i} U^*_{I \alpha i} \sin^2 \left(1.27 \frac{m^2_{
I\,j} - m^2_{I\,i}}{E} L \right)\;,
\end{equation}

\ni where $ E = \sqrt{p^2_i + m^2_{I\,i}}$ ($i = 1,2,3$).

 If the matrix $\widehat{m}^{(L)}_s $ happens to be nearly diagonal and has 
nearly degenerate eigenvalues: $ m^{(L)}_{s \alpha \beta} \simeq \delta_{\alpha
\beta} m^{(L)}_s $ , then


\begin{equation}
m_{I \alpha \beta} \simeq - \sum_\gamma \frac{\mu^{(L)}_{\alpha
\gamma}\mu^{(L) }_{\gamma \beta}}{m^{(L)}_s}\;,\; m_{II \alpha \beta}
\simeq \delta_{\alpha \beta} m^{(L)}_s
\end{equation}


\ni and Eqs. (20) give

\begin{equation}
m_{I i} \simeq - \frac{\mu^{(L)\,2}_i}{m^{(L)}_s}\;,\; m_{II i} \simeq  
m^{(L)}_s \;,
\end{equation}

\ni where the eigenvalues $\mu^{(L)}_i $ of $\widehat{\mu}^{(L)} = (\mu^{(L)
}_{\alpha \beta})$ are produced with the use of $\widehat{U}_I $:

\begin{equation}
\delta_{ij} \mu^{(L)}_i  = \sum_{\alpha \beta} U^*_{I\,\alpha\,i} U_{I\,\beta\,
j} \mu^{(L)}_{\alpha \beta}\;.
\end{equation}

\ni In this case, $\nu_{II \alpha}$ ($\alpha = e\,,\,\mu\,,\,\tau $) are 
(approximate) neutrino heavy mass eigenstates decoupled from $\nu_{I i}$ ($i = 
1,2,3 $).

 Concluding, the lefthanded see--saw, constructed in this note,
implies both the smallness of active--neutrino masses and decoupling
of heavy passive neutrinos, similarly to the situation in the case of
conventional see--saw. However, introducing as heavy passive neutrinos
the lefthanded sterile neutr%
inos $\nu_{\alpha L}^{(s)}$ in place of
the conventional righthanded neutrinos $\nu_{\alpha R}$ ($\alpha =
e\,,\,\mu\,,\,\tau $), the new see--saw mechanism, when spoiling the
chiral left--right pattern of original \SM for neutrinos, does it in a
way different from the conventional see--saw mechanism. Recall that in
the lefthanded see--saw the righthanded neutrinos continue to be
completely absent.

Note finally that in the case of lefthanded see--saw the active and
sterile ne%
utrinos, as practically unmixed, cannot oscillate into
each other. They could, if in Eq. (2) in place of the inequality (10)
the relation

\begin{equation}
\mid m^{(L)} - m^{(L)}_s \mid \ll \mu^{(L)}
\end{equation}

\ni were conjectured, leading to their nearly maximal mixing. This mechanism, 
however, being a formal analogy of the conventional pseudo--Dirac case [4], 
would not imply automatically small neutrino masses, since then $ m_{I,II} 
\simeq (1/2)(m^{(L)} + m^{(L)}_s) \mp \mu^{(L)}$ from Eq. (9). The smallness 
of $ m_I $ would require $(1/2)(m^{(L)} + m^{(L)}_s) \simeq \mu^{(L)}$.

 I am indebted to Stefan Pokorski for several helpful discussions.

\vfill\eject               

~~~~
\vspace{1.0cm}

{\centerline{\bf References}}

\vspace{0.3cm}

{\everypar={\hangindent=0.5truecm}
\parindent=0pt\frenchspacing

{\everypar={\hangindent=0.5truecm}
\parindent=0pt\frenchspacing

~1.~M.~Gell--Mann, P.~Ramond and R.~Slansky, in {\it Supergravity }, ed. 
P.~van Nieuwenhuizen and D.Z.~Freedman, North--Holland, Amsterdam, 1979;
T.~Yanagida, in {\it Proc. of the Workshop on the Unified Theory of the 
Baryon Number in the Universe}, ed. O.~Sawada and A.~Sugamoto, KEK report 
No. 79--18, Tsukuba, Japan, 1979; see also R.~Mohapatra and G.~Senjanovic,
{\it Phys. Rev. Lett.} {\bf 44}, 912 (1980).

\vspace{0.15cm}

~2.~R.D.~Peccei, hep--ph/9906509.

\vspace{0.15cm}

~3.~G.~Altarelli and F.Feruglio, CERN--TH/99--129 + DFPD--99/TH/21, hep--ph/%
9905536.

\vspace{0.15cm}

~4.~See {\it e.g.} W. Kr\'{o}likowski, hep--ph/9910308.

\vfill\eject

\end{document}